\documentclass[preprint,12pt]{elsarticle}
\usepackage{balance}
\usepackage{graphicx}
\usepackage{color}

\usepackage{hyperref}

\usepackage{amsthm}

\theoremstyle{definition}

\theoremstyle{remark}

\usepackage{listings}
\usepackage{eiffel}
\usepackage{boogie}

\usepackage{amsmath,amssymb}
\usepackage{wasysym}
\usepackage{enumerate}
\usepackage{color}   
\usepackage{amssymb}  

\begin{document}
\begin{frontmatter}
\title{AutoReq: expressing and verifying requirements for control systems}
%
%
\author[inno,ups]{Alexandr Naumchev}
\ead{a.naumchev@innopolis.ru}
\author[inno,poli,ups]{Bertrand Meyer}
\ead{b.meyer@innopolis.ru}
\author[inno]{Manuel Mazzara}
\ead{m.mazzara@innopolis.ru}
\author[ups]{Florian Galinier}
\ead{florian.galinier@irit.fr}
\author[ups]{Jean-Michel Bruel}
\ead{bruel@irit.fr}
\author[ups]{Sophie Ebersold}
\ead{ebersold@irit.fr}

\address[inno]{Innopolis University, Innopolis, Russia}
\address[ups]{Toulouse University, Toulouse, France}
\address[poli]{Politecnico di Milano, Milan, Italy}
\begin{abstract}
The considerable effort of writing requirements is only worthwhile if the result meets two conditions: the requirements reflect stakeholders' needs, and the implementation satisfies them.
In usual approaches, the use of different notations for requirements (often natural language) and implementations (a programming language) makes both conditions elusive.
AutoReq, presented in this article, takes a different approach to both the writing of requirements and their verification.
Applying the approach to a well-documented example, a landing gear system, allowed for a mechanical proof of consistency and uncovered an error in a published discussion of the problem.
\end{abstract}


\end{frontmatter}

\section{Overview and Main Results} \label{sec:overview}

A key determinant of software quality is the quality of requirements.
Inconsistent or incomplete understanding of the requirements can lead to catastrophic results.
This article presents a tool-supported method, AutoReq, for producing verified requirements, with applications to control systems.
It illustrates it on a standard case study, an airplane Landing Gear System (LGS).
The goal is to obtain requirements of high quality:
\begin{itemize}
	\item Easy to write.
	\item Clear and explainable to domain experts.
	\item Amenable to change.
	\item Supporting traceability through close connections to later development steps, particularly implementation.
	\item Amenable to mechanical verification and validation.
\end{itemize}

As the last point indicates, AutoReq includes techniques for not only expressing requirements but also verifying their consistency.
The LGS case study illustrated the effectiveness of such verification by uncovering a significant error in a previous description of this often-studied example (\autoref{sec:sec:error}).

AutoReq takes natural language requirements and environment assumptions as an input and converts them into a format having the above properties.
The new format relies on a programming language with contracts.
This viewpoint brings one of the biggest advantages of AutoReq -- it makes the requirements verifiable both against the underlying assumptions and future candidate implementations, while maintaining their readability through natural language comments on the code.
The present work takes the natural language statements from the LGS case study and translates them to seamless statements, readable and verifiable.
The ASM treatment of the case study \cite{arcaini2017rigorous} provides the candidate implementation -- an executable ASM specification \cite{gurevich1994evolving} of the system.
This by no means implies applicability of AutoReq to ASMs only.
The approach applies to any candidate implementation that follows the small step semantics of ASMs.
More precisely, the implementation should run in an infinite loop polling the system environment's state and sending appropriate control signals.
To the best of our knowledge, most control systems' implementations follow this approach.

The method of expressing requirements does not introduce any new formalism but instead relies on a standard programming language, Eiffel, using mechanisms of Design by Contract (DbC) \cite{meyer1992applying} to state semantic constraints.
While DbC relies on Hoare logic \cite{hoare1969axiomatic}, which at first sight does not cover temporal and timing properties essential to the specification of control systems, we show that it is, in fact, possible and even simple to express such properties in the DbC framework. 

The verification part relies on an existing tool, associated with the programming language: AutoProof \cite{tschannen2015autoproof}, a program proving framework, which can verify the temporal and timing properties expressed in the DbC framework.
Applying it to LGS automatically and unexpectedly uncovered the error. Hoped-for advantages include:
\begin{itemize}
	\item Expressiveness: requirements benefit both from the expressive power of declarative assertions and from that of imperative instructions.
	\item Ease of learning: anyone familiar with programming languages has nothing new to learn.
	\item Continuity with the rest of the development cycle: design and implementation may rely on the same formalism, avoiding the impedance mismatches that arise from the use of different formalisms, and facilitating change.
	\item Precision: formal specifications (contracts) cover the precise semantics of the system and its environment.
	\item Existing tools, as available in modern IDEs, that support the requirements process: a compiler for a typed language performs many checks that are as useful for requirements as for code.
\end{itemize}

The present work, while not claiming to have fully reached these ambitious goals, makes the following contributions:
\begin{itemize}
	\item The outline of a general method for requirements engineering with application to control systems.
	\item The use of a programming language as an effective mechanism for requirements specification.
	\item A precisely defined concept of \textit{verifying requirements} for control systems (complementing the usual concept of verifying programs).
	This idea originates from \cite{naumchev2017seamless}.
	\item A translation scheme from temporal and timing properties to simpler Hoare logic properties (essentially, first-order predicates on states) as traditionally used in Design by Contract.
	\item A simple way to combine \textit{environment} and \textit{machine} aspects (the two components of requirements in the well-known Jackson-Zave approach).
	\item A direct mapping of these \textit{requirements} concepts into well-known \textit{verification} concepts, \textit{assume} and \textit{assert}.
	\item The demonstration that it is possible to use an existing \textit{program} prover to verify requirements.
\end{itemize}

\autoref{sec:the_importance_of_verifying_requirements} discusses consequences of poor requirements.
\autoref{sec:landing_gear_example} presents LGS.
\autoref{sec:writing_requirements} describes the methodology: how to specify and verify requirements.
\autoref{sec:structuring_an_embedded_system_specification} shows how to translate common requirements patterns (originally expressed through temporal logic, timing constraints or Abstract State Machines) into a form suitable for AutoReq.
\autoref{sec:applying} sketches the method’s application to the case study, including an analysis of the uncovered error.
\autoref{sec:related_work} discusses related work, and \autoref{sec:conclusions} discusses limitations and future work.

\section{The importance of verifying requirements} \label{sec:the_importance_of_verifying_requirements}

Control systems in aerospace, transportation, and other mission-critical areas raise tough reliability demands.
Ensuring reliability begins with the quality of requirements: the best implementation is useless if the requirements are inconsistent or do not reflect needs.
Requirements for software deserve as much scrutiny as other artifacts such as code, designs, and tests.

The literature contains many examples of software disasters arising from requirements problems of two kinds:
\begin{itemize}
	\item In the requirements themselves: inconsistencies, incompleteness, inadequate reflection of stakeholders' needs.
	\item In their relationship to other tasks: design, implementation etc. may wrongly understand, implement or update them.
\end{itemize}

Examples of the first kind include \cite{lake2010epic}:
\begin{itemize}
	\item The year 2000, National Cancer Institute, Panama City: patients undergoing radiation therapy get wrong doses because of a software miscalculation.
	\item In 1996, Ariane 5 maiden flight fails from flight computer's code crash, out of an uncaught arithmetic exception, in code that was reused from Ariane 4 but relied on assumptions that no longer hold in the new technology.
	\item In 1990, a bug in software for AT\&T's \#4ESS long-distance switches crashes computers upon receipt of a specific message sent out by neighbors when recovering from a crash.
\end{itemize}

Analysis of these examples suggests that the problem lies in part from the use of different methods and of different notations for requirements and other tasks such as implementation.
This observation is a basis for the \textit{seamless} approach (\cite{meyer1997object}, \cite{walden1994seamless}, \cite{meyer2013multirequirements}, \cite{naumchev2017seamless}, following ideas in \cite{rumbaugh1991object}), which this article applies by using a single notation throughout.

Examples of the second kind include \cite{van2009requirements}:
\begin{itemize}
	\item London underground system: several cases \cite{neumann1994computer} of passenger deaths from doors opening or closing unexpectedly, without an alarm notification being sent to the train driver.
	\item An aerospace project \cite{hooks2001customer} where 49\% of requirements errors were due to incorrect facts about the problem world.
	\item An inadequate assumption about the environment of the flight guidance system, which may have contributed to the crash of a Boeing 757 in Cali \cite{modugno1997integrated}.
	Location information for the pilot to extend the flap arrived late, causing the guidance software to send the plane into a mountain.
\end{itemize}

These examples and others in the literature illustrate the importance of \textit{verifying} requirements.
We will see that it is possible to apply to requirements both the concept of verification, as commonly applied to code, and modern proof-oriented verification tools devised initially for code.

\section{The Landing Gear System} \label{sec:landing_gear_example}

To illustrate AutoReq, this article will use, rather than examples of the authors' own making, the LGS \cite{boniol2014landing}, probably the most widely discussed case study in recent control systems literature, e.g. \cite{su2017aircraft}, \cite{arcaini2017rigorous}, \cite{dhaussy2014context}, \cite{hansen2017validation}, \cite{mammar2017modeling}, \cite{berthomieu2014model}, \cite{banach2014landing}.

\begin{figure}
	\begin{center}
		\includegraphics[scale=0.6]{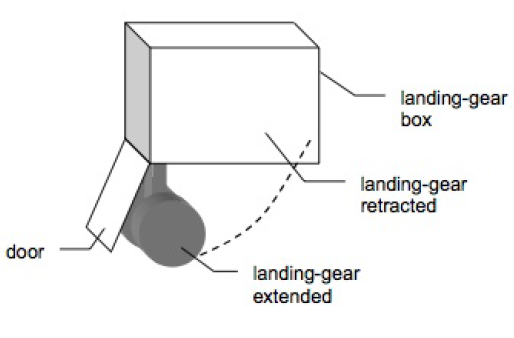}
	\end{center}
	\caption{Landing set (from Boniol et al. \cite{boniol2014landing}).}
	\label{fig:lgs}
\end{figure}

The Landing Gear System physically consists of the landing set, a gear box that stores the gear in the retracted position, and a door attached to the box (\autoref{fig:lgs}).
A digital controller independently actuates the door and the gear.
The controller initiates either gear extension or gear retraction depending on the current position of a handle in the cockpit.
The task is to program the controller so that it sends the correct signals to the door's and the gear's actuators.

The discussion will restrict itself to the system's \textit{normal mode} (there is also a \textit{failure mode}).
The defining properties are the following:
\begin{description}
	\item[$R_{11}bis$:]\ When the landing gear handle has been pushed down and stays down, then eventually the gear will be seen extended and the doors will be seen closed.
	We interpret this requirement in LTL as $\square (\square handle\_down \implies \Diamond (gear\_extended \land door\_closed))$ where $\square$ stands for the \textit{always} temporal operator, and $\Diamond$ stands for the \textit{eventually} temporal operator.
	\item[$R_{12}bis$:]\ When the landing gear handle has been pulled up and stays up, then eventually the gears will be seen retracted and the doors will be seen closed. We interpret this requirement in LTL as $\square (\square handle\_up \implies \Diamond (gear\_up \land door\_closed))$.
	\item[$R_{21}$:]\ When the landing gear handle remains in the \textit{down} position, then retraction sequence is not observed.
	We interpret this requirement in LTL as $\square (handle\_down \implies \Circle \lnot gear\_retracting)$ where $\Circle$ stands for the \textit{next} temporal operator.
	\item[$R_{22}$:]\ When the landing gear handle remains in the \textit{up} position, then outgoing sequence is not observed.
	We interpret this requirement as $\square (handle\_up \implies \Circle \lnot gear\_extending)$.
\end{description}

We will work not directly from the original description of the LGS but from one of the most interesting treatments of case study \cite{arcaini2017rigorous}, which uses the abstract state machine (ASM) approach and applies a process of successive refinements:
\begin{enumerate}
	\item Start with a \textit{ground model} covering a subset of the requirements.
	\item Model-check it.
	\item Repeatedly extend (refine) it with more properties of the system, proving the correctness of each refinement.
\end{enumerate}

The AutoReq specification discussed in the next sections starts from the ASM ground model.
Some of its features are a consequence of this choice:
\begin{itemize}
	\item It only accounts for properties specified in the first of the successive models in \cite{arcaini2017rigorous}.
	\item As already noted, it only covers \textit{normal mode}.
	\item Like the ASM model, it assumes that the only \textit{environment-controlled} \textit{machine-visible} phenomenon is the pilot's handle \cite{jackson1995deriving}.
	In the failure mode, there might be others.
	\item It takes over from the ASM model such instructions as $gears := RETRACTED$ which posit that the control system has a way to send the gear directly to the retracted position.
	This assumption is acceptable at the modeling level but not necessarily true in the actual LGS system.
	\item The ASM-to-Eiffel translation scheme (\autoref{sec:sec:translating_asm_properties}) ensures preservation of the one-step semantics of ASM.
\end{itemize}

\section{Requirements methodology} \label{sec:writing_requirements}

AutoReq builds on the ideas of seamless development \cite{meyer1997object}, \cite{walden1994seamless}, multirequirements \cite{meyer2013multirequirements} and seamless requirements \cite{naumchev2017seamless}.
The new focus is on requirements verification and reuse of previous requirements through a routine call mechanism.
We examine in turn how to specify and reuse requirements and environment assumptions (\autoref{sec:sec:specifying_requirements}), and what it means to verify them (\autoref{sec:sec:verifying_requirements}).

\subsection{Specifying requirements} \label{sec:sec:specifying_requirements}

Specifications in AutoReq, often in practice translated from a document in structured natural language, take the form of contracted Eiffel routines with natural-language comments.
These routines are further consumed by:
\begin{itemize}
	\item The verification tool. Since the routines coming out of the translation process are equipped with contracts, they may be formally verified by a Hoare logic based prover.	
	\item Possible implementers of the system. The combination of a programming language and natural language helps developers, who will use the same programming language for implementation, understand the requirements. The contracts state the semantics.
\end{itemize}

Previous publications (\cite{meyer2013multirequirements, naumchev2017seamless}) explain the reasons for choosing this mixed notation: unity of software construction and verification, unity of functional requirements and code, use of complementary notations geared towards different stakeholders.

Additional properties are specific to control systems: 
\begin{itemize}
	\item Specification of temporal assumptions and requirements.
	\item Specification of timing assumptions and requirements.
	\item Reuse of assumptions and requirements in stating new ones.
\end{itemize}

The basic notation is Eiffel. All the examples have been processed by the EiffelStudio IDE \cite{EiffelCommunity}, compiled, and processed by the AutoProof verification environment.
The interest of compilation is not in the generated code, since at this stage the Eiffel texts represent requirements only, but in the many consistency controls, such as type checking, of a modern compiler.

The requirements can and do take advantage of object-oriented mechanisms such as classes, inheritance and genericity.

There is sometimes an instinctive resistance to using a programming language for requirements, out of the fear of losing the fundamental difference between the goals of the two steps: programming languages normally serve for implementation, while requirements should be descriptive.
The AutoReq approach, however, uses the programming language not for implementation but for specification, restricting itself to requirements patterns discussed next.
The imperative nature of these patterns does not detract from this goal; empirical evidence indeed suggests \cite{fahland2009declarative} that operational reasoning works well not just for programmers but for other requirements stakeholders.
An added benefit is the availability of program verification tools, which the approach of this article channels towards the goal of verifying requirements.

For this verification goal, there seems to be a mismatch between the standard properties that program verification tools address and the needs of control systems.
Program verification generally relies on Hoare logic properties as embodied in Eiffel's Design by Contract: properties of the program state (or, for postconditions, two states).
The specification of control systems generally relies on temporal and timing requirements, involving properties of an arbitrary number of (future) states of the system.
A contribution of this work is to resolve the mismatch, using the programming language to emulate temporal and timing properties, through schemes described in \autoref{sec:structuring_an_embedded_system_specification}.

\subsection{Verifying requirements} \label{sec:sec:verifying_requirements}

Verification of AutoReq requirements relies on AutoProof \cite{tschannen2015autoproof}, the prover of contracted Eiffel programs.
AutoProof is a Hoare logic \cite{hoare1969axiomatic} based verifier that follows \textit{semantic collaboration} \cite{polikarpova2014flexible} -- a specification and verification methodology adapting Hoare logic to specific needs of object-oriented programming.
The verification unit of AutoProof is feature with contracts.
AutoReq assumptions and requirements take the form of such features, with natural language comments for better readability, to enable their direct verification with AutoProof.

Contracts for verification with AutoProof may be modular -- visible to the feature's callers, and non-modular -- visible only in the feature's implementation.
Modular contracts take the following forms:
\begin{itemize}
	\item \textit{Precondition} imposes obligations on the feature's callers and benefits the callees' implementation.
	\item \textit{Postcondition} guarantees benefits to the callers and imposes obligations on the callees' implementation.
\end{itemize}

Non-modular contracts take the following forms, going back at least as far as ESC-Java \cite{cok2004esc}:
\begin{itemize}
	\item \textbf{assume} X \textbf{end} \textit{allows} the verification to take advantage, at the given program point, of property X, adding X to the set of properties that the prover \textit{may} use (assumption).
	\item \textbf{assert} X \textbf{end} \textit{requires} the verification to establish X before going beyond the program point, adding X to the set of properties that the prover must prove (proof obligation).
\end{itemize} 

Both \textit{precondition} and \textit{assume} contracts add information to verifying the \textit{postcondition} and \textit{assert} contracts, but preconditions impose verification obligations on their own: they have to hold whenever the respective features are called.
AutoReq requirements take the form of features with non-modular contracts because of their fundamental connection with the core requirements engineering terminology, as discussed further.
From the purely technological perspective, AutoReq depends on the ability of AutoProof to inline callees' non-modular contracts into the callers' code.

As noted in the introduction, many software errors are requirements errors.
To avoid inconsistencies, AutoReq specifications include formal properties which can be submitted to proof tools for verification.
Jackson \& Zave's seminal work (\cite{jackson1995deriving}, also van Lamsweerde \cite{van2009requirements}), introduced a fundamental division of these properties:
\begin{itemize}
	\item \textbf{Environment} (or \textbf{domain}) assumptions characterize the context in which the system must operate.
	The development team has no influence on them.
	
	\item \textbf{Machine} (or \textbf{system}) properties characterize what the system must do.
	It is the job of the development team to work on them.
\end{itemize}

Although each of these two distinctions is well-known and widely used in the corresponding sub-community of software engineering, respectively requirements and formal verification, the existing literature does not, to our knowledge, connect them.
The present work, covering both requirements and verification concepts, unifies them into a single distinction:
\begin{itemize}
	\item \textbf{assume} E \textbf{end} specifies an \textit{environment assumption} E.
	\item \textbf{assert} E \textbf{end} specifies a \textit{machine property} E.
\end{itemize}
Verifying requirements in AutoReq simply means proving that all \textbf{\textit{assert}} hold, being permitted to take \textbf{\textit{assume}} for granted.

Notational convention: the above notations are for presentation.
The actual texts verified through the process reported in the next sections use the following standard Eiffel equivalents:
\begin{itemize}
	\item For \textit{\textbf{assert} X \textbf{end}}, the notation in the actual Eiffel texts is \textit{\textbf{check} X \textbf{end}} (\textit{\textbf{check}} is a standard part of Eiffel's Design by Contract mechanism).
	\item For \textit{\textbf{assume} X \textbf{end}}, the Eiffel notation is \textit{\textbf{check} assume: X \textbf{end}}.
	The \textit{assume} tag is a standard part of the notation for programs to be verified by AutoProof.
	\textit{\textbf{old} e}, in a routine body, denotes the value of an expression \textit{e} on routine entry.
\end{itemize}

The only difference with verifying programs comes from the elements that appear between these assertions: in program verification, they may include any instructions; in requirements verification, we only permit patterns discussed below \autoref{sec:sec:overall_program_structure}.
In addition, specifications include timing properties, using the translation into classic assertions described in \autoref{sec:sec:translating_temporal_properties} and \autoref{sec:sec:translating_timing_properties}.

Formal methods and notations are essential for one of the goals of this work (precision/completeness, see \autoref{sec:overview}), but non-technical stakeholders sometimes find them cryptic at first sight, hampering other goals such as readability and ease of use.
The \textit{multirequirements} approach \cite{meyer2013multirequirements}, which this article extends, addresses the problem by using complementary views, kept consistent, in various notations: formal (such as Eiffel or a specification language), graphical (such as UML) and textual (such as English).
In line with this general idea, AutoReq specifications rely on systematic commenting conventions (somewhat in the style of Knuth's \textit{literate programming} \cite{knuth1984literate}).
A typical example from the specification in the next section is
\begin{lstlisting}
-- Assume the system
run_in_normal_mode
\end{lstlisting}

The second line is formal; the comment in the first line puts it in context.
Such seemingly informal comments follow precise rules.
For non-expert users, and for the present discussion, it is enough to treat them as natural-language explanations.

\section{Structuring a control system specification} \label{sec:structuring_an_embedded_system_specification}

The mechanisms of the preceding section enable us to write the requirements for control systems and verify them.
Such specifications will follow standard patterns:
\begin{itemize}
	\item Overall structure of programs that model control systems (\autoref{sec:sec:overall_program_structure}).
	\item Translation rules for temporal properties (\autoref{sec:sec:translating_temporal_properties}).
	\item Translation rules for timing properties (\autoref{sec:sec:translating_timing_properties}).
	\item Translation rules for ASM properties (\autoref{sec:sec:translating_asm_properties}).
\end{itemize}

These schemes and translation patterns are fundamental to the methodology because they govern the use of the programming language.
While the methodology relies on a programming language for expressing requirements, it does not use its full power, since some of its mechanisms are only relevant for programs.
Programming language texts expressing requirements stick to the language subset relevant to this goal.

The translation schemes of \autoref{sec:sec:translating_temporal_properties}, \autoref{sec:sec:translating_timing_properties} and \autoref{sec:sec:translating_asm_properties} guarantee that their output will conform to these patterns.
A goal for future work (\autoref{sec:conclusions}) is to formalize the input languages, timed temporal logic and ASM, and turn the translation patterns into formal rules and automatic translation tools.

Pending such formalization, we did not for now address the soundness of the translation.

\subsection{Representing control systems} \label{sec:sec:overall_program_structure}

A control system is typically (unlike most sequential programs) repeating and non-terminating.
AutoReq correspondingly uses programs of the form \textit{\textbf{from} \textbf{until} \textbf{False} \textbf{loop} main \textbf{end}}.
The task of the requirements is then to specify \textit{main}.

The translation uses four patterns that look like Eiffel features with non-modular (\textit{assume} and \textit{assert}) contracts.
These patterns are not part of AutoProof, but they serve as blueprints for features that AutoProof can verify.
\textit{P1} and \textit{P2} (\autoref{sec:sec:translating_temporal_properties}) are time-independent (although \textit{temporal} in the sense of temporal logic).
\textit{P3} and \textit{P4} (\autoref{sec:sec:translating_timing_properties}) take timing into account.
These cases suffice for the examples addressed with AutoReq so far.
Translation schemes are possible for more general LTL/CTL/TPTL schemes if the need arises in the future.

The patterns use the Jackson-Zave distinction (\autoref{sec:sec:verifying_requirements}) between describing an environment assumption and prescribing an expected system (machine) property.
Specifically: \textit{P1} and \textit{P3} correspond to environment assumptions (respectively time-independent and timed); \textit{P2} and \textit{P4} correspond to system obligations (with the same distinction).
The Eiffel translations accordingly use \textbf{\textit{assume}} for \textit{P1} and \textit{P3} and \textbf{\textit{assert}} for \textit{P2} and \textit{P4}.
When asked to verify an AutoReq requirement, AutoProof tries to infer the \textit{assert} statements by simulating an execution of the requirement's body to a state satisfying the \textit{assume} statements.
\begin{figure}
\begin{center}
\begin{tabular}{|c|c|c|}
	\hline 
	& \textbf{Temporal Properties} & \textbf{Timing Properties} \\ 
	\hline 
	\textbf{Environment Assumptions} & P1 & P3 \\ 
	\hline 
	\textbf{System Obligations} & P2 & P4 \\ 
	\hline 
\end{tabular} 
\end{center}
\caption{The map of AutoReq translation patterns.}
\label{tab:patterns}
\end{figure}
\autoref{tab:patterns} maps the patterns according to the taxonomy of system properties used in the present article.

\subsection{Translating temporal properties} \label{sec:sec:translating_temporal_properties}

In the control systems world, the starting point for requirements is often a description expressed in a temporal logic, usually LTL \cite{pnueli1977temporal}, CTL \cite{ben1983temporal}, or a timed variant such as propositional temporal logic (TPTL \cite{alur1994really}).
Even if not using a specific formalism, they often state temporal properties such as \textit{all future system states must satisfy a given condition} or \textit{some future state must satisfy a given condition}.
The LGS properties given in \autoref{sec:landing_gear_example} are an example.

\begin{itemize}
	\item \textbf{P1} (environment assumption)\\
	Consider the system running in mode $cs$ under assumption $c$.
	The LTL formulation is $\square (c \land cs)$. 
	\item \textbf{P2} (system obligation)\\
	The system running in mode $cs$ should immediately meet property $p$.
	The LTL formulation is $\square(cs \implies \Circle p)$.
	This property constrains the system to maintain response $p$ whenever stimulus $cs$ holds.
\end{itemize}

The translation scheme for \textit{P1} is:
\begin{lstlisting}
-- Assume the system
run_under_condition_c
  do
    assume
      c
    end
    main_under_conditions_cs
  end
\end{lstlisting}
where \textit{main\_under\_conditions\_cs} is of the form \textit{P1} or \textit{P3}.
The \textit{run\_under\_conditions} routine should be used instead of the original \textit{main} in all requirements that talk about the system operating in mode \textit{c}.
This pattern may be useful for encoding $\square c$ in properties of the form $\square (\square c \implies \Diamond d)$.

The translation scheme for \textit{P2} is:
\begin{lstlisting}
-- Require the system to
immediately_meet_property_p
  do
    main_under_conditions_cs
    assert
      p
    end
  end
\end{lstlisting}
where \textit{main\_under\_conditions\_cs} is of the form \textit{P1} or \textit{P3}.

\subsection{Translating timing properties} \label{sec:sec:translating_timing_properties}

Although not all approaches to requirements take time into account, timing requirements, such as \textit{the response time must not exceed 1 second}, are essential to the proper specification and implementation of control systems. AutoReq recognizes the following timing-related patterns:
\begin{itemize}
	\item \textbf{P3} (environment assumption)\\
	Assume the system running in mode $cs$ spends $t$ time units to meet property $p$.
	The TPTL formulation is $\square x. ((cs \land \lnot p) \implies \Circle y.(p \implies y = x + t))$.
	$x.$ and $y.$ record the current time of corresponding states \cite{alur1994really}.
	\item \textbf{P4} (system obligation)\\
	The system running in mode $cs$ should spend no more than $t$ time units to meet property $p$.	In TPTL: $\square x.(\square cs \implies \Diamond y.(p \land y \leq x + t))$.
\end{itemize}

The translation scheme for \textit{P3} is:
\begin{lstlisting}
-- Assume it takes t time units to take the system
from_not_p_to_p:
  do
    main_under_conditions_cs
    if (not old p and p) then
      duration := duration + t
    end
  end
\end{lstlisting}

The technique for timing system obligations of the \textit{P4} form differs from the others by using loops as the core mechanism:
\begin{lstlisting}
-- Require that
meeting_p_under_persistent_conditions_cs
-- never takes more than t time units:
  do
    from
      main_under_conditions_cs
    until
      p or (duration - old duration) > t
    loop
      main_under_conditions_cs
    end
    assert
      p and (duration - old duration) <= t
    end 
  end
\end{lstlisting}
where \textit{main\_under\_conditions\_cs} is of the form \textit{P1} or \textit{P3}.
The \textit{(duration -- \textbf{old} duration) \textgreater t} exit timeout condition ensures termination of the loop, and assertion \textit{(duration -- \textbf{old} duration) $\le$ t} checks that the timeout condition has not been reached.

The technique for handling the timing-related patterns relies on an integer, non-decreasing auxiliary variable \textit{duration}.
It has the same role as $x$ and $y$ in the TPTL formulations.
The \textit{duration} variable is part of the AutoReq approach -- not a predefined variable nor part of AutoProof.
It does not play a role in the actual execution of the system but caters to static reasoning about the system's timing properties.
The \textit{from\_not\_p\_to\_p} routine updates the value of \textit{duration} instead of using \textbf{\textit{assume}}, which would lead to a contradiction: the prover would detect that the variable was not, in fact, updated, and would infer \textbf{\textit{False}} from assuming the opposite.

\subsection{Translating ASM properties} \label{sec:sec:translating_asm_properties}

Abstract State Machines \cite{gurevich1994evolving}, \cite{borger2012abstract}, \cite{borger2018modeling} are a commonly used specification formalism for control systems, and the treatment of the LGS case study in \cite{arcaini2017rigorous} served as a starting point for this article's own treatment of the example.
The present work does not formally prove soundness of the ASM-to-Eiffel translation.
The decision to work with the ASM treatment was motivated by the general ASM specifications' executability: fundamentally, they are verifiable abstractions of infinitely running control software.
Such software may be implemented in a general-purpose programming language, and the present article demonstrates that such a language may serve as a verifiable abstraction of itself, in the presence of a program prover.

Below comes the ASM-to-Eiffel translation scheme.
The translation scheme omits the nondeterministic version of the ASM formalism.
The original work \cite{gurevich1994evolving} presents \textit{``Nondeterministic Sequential Algebras''} as an extension to the basic model. 
As \autoref{sec:overview} explains, the ASM formalism serves as an implementation language example in the present discussion of AutoReq, with no intent of covering every aspect of ASMs.
Nondeterministic updates seem to be inappropriate for implementing mission- and life-critical systems, such as the LGS, and control systems in general.
Every possible environment's state should be predictably handled in such systems.
The ASM treatment of the LGS, for example, does not use nondeterminism.

A basic ASM specification is a collection of rules taking one of three forms \cite{gurevich2000sequential}: assignment, do-in-parallel and conditional.
An ASM \textit{assignment} reads:
\begin{equation}
f(t_1,..,t_j):=t_0
\label{eq:asm_assignment}
\end{equation}
The semantics is: update the current content of location $\lambda = (f,(a_1,..,a_j))$, where $a_{i: \{1..j\}}$ are values referenced by $t_{i: \{1..j\}}$, with the value referenced by $t_0$. 

The Eiffel representation for an ASM location is an attribute (field) of the class; the representation for a location update is an attribute assignment.

The ASM \textit{do-in-parallel} operator applies several assignments in one step.
Eiffel offers no native support for do-in-parallel, but it can emulate one sequentially without changing the behavior.
The following example gives intuition behind the translation idea:
\begin{equation}
a,b := max (a-b, b), min (a-b,b)
\label{eq:asm_assignment_example}
\end{equation}
The instruction in \autoref{eq:asm_assignment_example}, when run infinitely, reaches the fixpoint in which $a$ contains the greatest common divisor of $a$ and $b$.
The Eiffel translation of this instruction is:
\begin{lstlisting}
	local
	  a_intermediate, b_intermediate: INTEGER
	do
	  a_intermediate := max (a-b, b)
	  b_intermediate := max (a-b, b)
	  a := a_intermediate
	  b := b_intermediate
	end
\end{lstlisting}
The generalization should be clear at this point: instead of updating directly the locations, introduce and update intermediate local variables, and then assign them to the locations.

The translation of an ASM \textit{conditional} (\textit{\textbf{if} t \textbf{then} R \textbf{else} Q}) is an Eiffel conditional instruction.

The ASM-to-Eiffel translation scheme scales out to the multiple classes case.
The translation overhead in this case consists of implementing assigner procedures for the supplier classes' attributes.
The assigner procedures will make it possible for the clients to update the suppliers' attributes while keeping them consistent.
The LGS example is simple enough to avoid the multiple classes case, which is why the present work never applies this translation rule.

\section{The Landing Gear System in AutoReq} \label{sec:applying}

Equipped with the AutoReq mechanisms as described, we can now see the core elements of the AutoReq specification of the LGS example.
The entire example is available in a public GitHub repository \cite{Naumchev2017LgsRepo}.

\subsection{Normal mode of execution} \label{sec:sec:normal_mode}

Execution runs in \textit{normal mode} if all the parameter values are in the expected ranges and meet the system invariant.
Application of the \textit{run\_under\_condition\_c} pattern results in the following Eiffel model of normal mode:
\begin{lstlisting}
-- Assume the system
run_in_normal_mode
  do
    -- the handle status range:
    assume
      handle_status = up_position or
      handle_status = down_position
    end
    -- the door status range:
    assume
      door_status = closed_position or
      door_status = opening_state or
      door_status = open_position or
      door_status = closing_state
    end
    -- the gear status range:
    assume
      gear_status = extended_position or
      gear_status = extending_state or
      gear_status = retracted_position or
      gear_status = retracting_state
    end
    -- the gear may extend or retract only with the door open:
    assume
      (gear_status = extending_state or gear_status = retracting_state)
        implies door_status = open_position
    end
    -- closed door assumes retracted or extended gear
    assume
      door_status = closed_position implies
        (gear_status = extended_position or gear_status = retracted_position)
    end
    main
  end
\end{lstlisting}
The first three \textit{\textbf{assume}} express that attribute values fall into specific ranges.
The last two express the LGS invariant.
Ranges, the invariant and the definition of normal mode come from the original.
\textit{run\_in\_normal\_mode} is a multiple application of the \textit{run\_under\_condition\_c} pattern (\autoref{sec:sec:translating_temporal_properties}).
It wraps around \textit{main} to make additional assumptions before calling it.

\subsection{Timing properties} \label{sec:sec:expressing_timing_constraints}

The ASM treatment of the LGS case study ignores timing properties stated in the original description.
For a practical system, timing is essential; an otherwise impeccable LGS that takes two hours to perform \textit{extend landing gear} would not be attractive.
We rely on AutoReq's timing mechanisms of the AutoReq methodology (\autoref{sec:sec:translating_timing_properties}) and the \textit{from\_not\_p\_to\_p} pattern (\autoref{sec:sec:translating_timing_properties}).
Timing values, e.g. 8 units for door closing, are for illustration only.
Each of the translations that follow are produced by applying the same pattern, which is why only the first translation is accompanied by a detailed explanation.

\begin{itemize}
	\item \textit{It takes 8 time units for the door to close.}
	Replacing \textit{p} with \textit{door\_status = closed\_position}, and \textit{t} with \textit{8} in the \textit{from\_not\_p\_to\_p} pattern yields:
	\begin{lstlisting}
	-- Assume it takes 8 time units to take the door
	from_open_to_closed -- position:
	  do
	    run_in_normal_mode
	    if (old door_status /= closed_position and
	       door_status = closed_position) then
	      duration := duration + 8
 	  end end
	\end{lstlisting}
	\item \textit{It takes 12 time units for the door to open}:
	\begin{lstlisting}
	--Assume it takes 12 time units to take the door
	from_closed_to_open -- position:
	  do
	    from_open_to_closed
	    if (old door_status /= open_position and
	       door_status = open_position) then
	      duration := duration + 12
	    end
	  end
	\end{lstlisting}
	\item \textit{It takes 10 time units for the gear to retract}:
	\begin{lstlisting}
	--Assume it takes 10 time units to take the gear
	from_extended_to_retracted -- position:
	  do
	    from_closed_to_open
	    if (old gear_status /= retracted_position and
	       gear_status = retracted_position) then
	      duration := duration + 10
	  end end
	\end{lstlisting}
	\item \textit{It takes 5 time units for the gear to extend}:
	\begin{lstlisting}
	-- Assume it takes 5 time units to take the gear
	from_retracted_to_extended -- position:
	  do
	    from_extended_to_retracted
	    if (old gear_status /= extended_position and
	       gear_status = extended_position) then
	      duration := duration + 5
	  end end
	\end{lstlisting}
\end{itemize}

\textit{from\_retracted\_to\_extended} will include all the previously stated \textbf{\textit{assume}} instructions together with \textit{main}.
\subsection{Baseline requirements} \label{sec:sec:baseline_requirements}

\autoref{sec:landing_gear_example} introduced a set of core LGS requirements, $R_{11}bis$ to $R_{22}$, which we now express in AutoReq.
$R_{11}bis$ and $R_{21}$ talk about the system running with the handle pushed down.
Application of the \textit{run\_under\_condition\_c} pattern (\autoref{sec:sec:translating_temporal_properties}) with \textit{handle\_status = down\_position} for \textit{c} results in the following routine to model the required mode of operation:
\begin{lstlisting}
-- Assume the system
run_with_handle_down
  do
    assume handle_status = down_position end
    from_retracted_to_extended
  end
\end{lstlisting}
\textit{run\_with\_handle\_down} is an application of the \textit{run\_under\_condition\_c} pattern (\autoref{sec:sec:translating_temporal_properties}).
It calls \textit{from\_retracted\_to\_extended} to include all assumptions so far.

Now that the execution mode with the handle pushed down is formally defined, it is possible to express the requirements in terms of it.
Property $R_{21}$ requires the controller to prevent retraction immediately whenever the handle is pushed down.
Application of the \textit{immediately\_meet\_property\_p} pattern (\autoref{sec:sec:translating_temporal_properties}) with \textit{gear\_status /= retracting\_state} for \textit{p} yields, for $R_{21}$:
\begin{lstlisting}
-- Require the system to
never_retract_with_handle_down
  do
    run_with_handle_down
    assert gear_status /= retracting_state end
  end
  -- known as R_{21} 
\end{lstlisting}

$R_{11}bis$ requires the system eventually to extend the gear and close the door if the handle stays down.
The absence of timing makes it unsuitable for the specification of control systems: we need to specify an upper bound on the time the system may spend on gear extension.
That bound is the sum of the maximal times for door closing, door opening and gear extension.
Under earlier assumptions, this value is 25.
Applying \textit{meeting\_p\_under\_persistent\_conditions\_cs} (\autoref{sec:sec:expressing_timing_constraints}) with \textit{gear\_status = extended\_position \textbf{and} door\_status = closed\_position} for \textit{p}, \textit{run\_with\_handle\_down} for \textit{main\_under\_conditions\_cs} and \textit{25} for \textit{t} turns $R_{11}bis$ into:
\begin{lstlisting}
-- Require that
extension_duration
-- never takes more than 25 time units:
  do
    from
      run_with_handle_down
    until
      (gear_status = extended_position and door_status = closed_position) or
      (duration - old duration) > 25
    loop
      run_with_handle_down
    end
    assert gear_status = extended_position end
    assert door_status = closed_position end
    assert (duration - old duration) <= 25 end
  end
  -- known as R_{11}bis 
\end{lstlisting}

Requirements $R_{12}bis$ and $R_{22}$ talk about the system running with the handle pulled up.
Application of \textit{run\_under\_condition\_c} (\autoref{sec:sec:translating_temporal_properties}) with \textit{handle\_status = up\_position} for \textit{c} yields:

\begin{lstlisting}
-- Assume the system
run_with_handle_up
  do
    assume
      handle_status = up_position
    end
    from_retracted_to_extended
  end
\end{lstlisting}

The rest of the requirements can rely on the specification of the execution mode with handle up, as we have now obtained.

$R_{22}$ requires the system to prevent immediate extension whenever the handle is pulled up.
Application of \textit{immediately\_meet\_property\_p} (\autoref{sec:sec:translating_temporal_properties}) with \textit{gear\_status /= extending\_state} for \textit{p} yields, for $R_{22}$:
\begin{lstlisting}
-- Require the system to
never_extend_with_handle_up
  do
    run_with_handle_up
    assert
      gear_status /= extending_state
    end
  end
  -- known as R_{22}
\end{lstlisting}

$R_{12}bis$ requires the system eventually to retract the gear and close the door if the handle stays up.
Like $R_{11}bis$, it does not include timing.
The upper bound for $R_{12}bis$ is the sum of the maximal times for door closing, door opening and gear extension, 30 from earlier assumptions.
Applying \textit{meeting\_p\_under\_persistent\_conditions\_cs} (\autoref{sec:sec:expressing_timing_constraints}) with \textit{gear\_status = retracted\_position \textbf{and} door\_status = closed\_position} for \textit{p}, with \textit{run\_with\_handle\_up} for \textit{main\_under\_conditions\_cs} and \textit{30} for \textit{t} yields:

\begin{lstlisting}
-- Require that
retraction_duration
-- never takes more than 30 time units:
  do
    from
      run_with_handle_up
    until
      (gear_status = retracted_position and door_status = closed_position) or
      (duration - old duration) > 30
    loop
      run_with_handle_up
    end
    assert
      gear_status = retracted_position and
      door_status = closed_position and
      (duration - old duration) <= 30
    end
  end
  -- known as R_{12}bis
\end{lstlisting}

\subsection{Complementary requirements} \label{sec:sec:complementary_requirements}

$R_{11}bis$ and $R_{12}bis$ talk about reaching a desired state under some conditions, but not about preserving it.
For example, even if the gear becomes extended and the door closed with the handle down, this situation must not change without the handle pulled up.
The following application of \textit{immediately\_meet\_property\_p} (\autoref{sec:sec:translating_temporal_properties}) with \textit{gear\_status = extended\_position \textbf{and} door\_status = closed\_position} for \textit{p} captures this property:
\begin{lstlisting}
-- Require the system to
keep_gear_extended_door_closed_with_handle_down
  do
    run_with_handle_down_gear_extended_door_closed
    assert
      gear_status = extended_position and
      door_status = closed_position
    end
  end
\end{lstlisting}
under the assumption that the doors are already closed, the gear is extended, and the handle is down.
Application of \textit{run\_under\_condition\_c} (\autoref{sec:sec:translating_temporal_properties}) with \textit{gear\_status = extended\_position \textbf{and} door\_status = closed\_position} for \textit{c} yields, for this assumption:
\begin{lstlisting}
-- Assume the system
run_with_handle_down_gear_extended_door_closed
  do
    assume
      gear_status = extended_position and
      door_status = closed_position      
    end
    run_with_handle_down
  end
\end{lstlisting}

The state with the gear retracted, the door closed and the handle pulled up should be stable without pushing the handle down.
The following application of \textit{immediately\_meet\_property\_p} (\autoref{sec:sec:translating_temporal_properties}) with \textit{gear\_status = retracted\_position \textbf{and} door\_status = closed\_position} for \textit{p} yields:
\begin{lstlisting}
-- Require the system to
keep_gear_retracted_door_closed_with_handle_up
  do
    run_with_handle_up_gear_retracted_door_closed
    assert
      gear_status = retracted_position and
      door_status = closed_position
    end
  end
\end{lstlisting}
under the assumption that the doors are already closed, the gear is retracted, and the handle is up.
Application of \textit{run\_under\_condition\_c} pattern (\autoref{sec:sec:translating_temporal_properties}) with \textit{gear\_status = retracted\_position \textbf{and} door\_status = closed\_position} for \textit{c} yields, for this assumption:
\begin{lstlisting}
-- Assume the system
run_with_handle_up_gear_retracted_door_closed
  do
    assume
      gear_status = retracted_position and
      door_status = closed_position
    end
    run_with_handle_up
  end
\end{lstlisting}

\subsection{An error in the ground model} \label{sec:sec:error}

Contracts do not just yield expressive power: they also make automatic verification possible in the AutoReq approach thanks to AutoProof.
One of the principal potential benefits would be to uncover errors in the requirements.

Our work on the LGS example shows that this benefit is not just a theoretical possibility.
Applying the AutoReq method and tools to the published ASM specification of the LGS system \cite{arcaini2017rigorous} uncovered an error.
The verification process applied the following sequence of steps.

\textbf{Start from the ASM specification.}
The language in which the ASM specification is expressed contains syntactic sugar in addition to the standard ASM operators.
The first step consisted of analyzing these additional constructs to understand how they should translate to Eiffel.

\textbf{Translate it into Eiffel.}
This step consisted of manual translation of the specification and the requirements to Eiffel.
One can find the original ASM specification in an online archive \cite{LgsAsmRepo}, inside the \textit{LandingGearSystemGround.asm} file.
File \textit{ground\_model.e} in the GitHub repository \cite{Naumchev2017LgsRepo} contains the result of the translation.

\textbf{Verify it with AutoProof.}
Note that AutoProof, by default, performs modular contract-based verification.
AutoReq specification techniques rely on \textbf{\textit{assume}} and \textbf{\textit{assert}} rather than traditional contracts.
These specification techniques require tuning AutoProof command-line options.
The GitHub repository \cite{Naumchev2017LgsRepo} with the Eiffel translation includes a \textit{readme} file that says in detail how to launch AutoProof.

\textbf{Identify the error.}
When AutoProof reports a verification failure, it does not point at its root cause.
The last step was devoted to identifying that cause.

The error uncovered by this procedure is subtle and revealing:
\textit{The specification does not meet the $R_{11}bis$ requirement, which states that pushing the handle down should lead to the gear extended and the door closed.}
Normally, when the crew pushes the LGS handle down, the controller should initiate the gear extension process.
Regardless of the initial system's state, this process should end up correctly -- so that in the end the gear is extended and the LGS latch is closed.

There exists, however, a state from which the erroneous ASM specification will not bring the system to the correct configuration.
This state corresponds to a situation in which the gear has just been retracted, the door is closing, and the crew decides to cancel retraction by pushing the handle down.
A correctly working system would cancel the retraction sequence and initiate gear extension.
\begin{figure}
	\centering
	\includegraphics[width=0.9\linewidth]{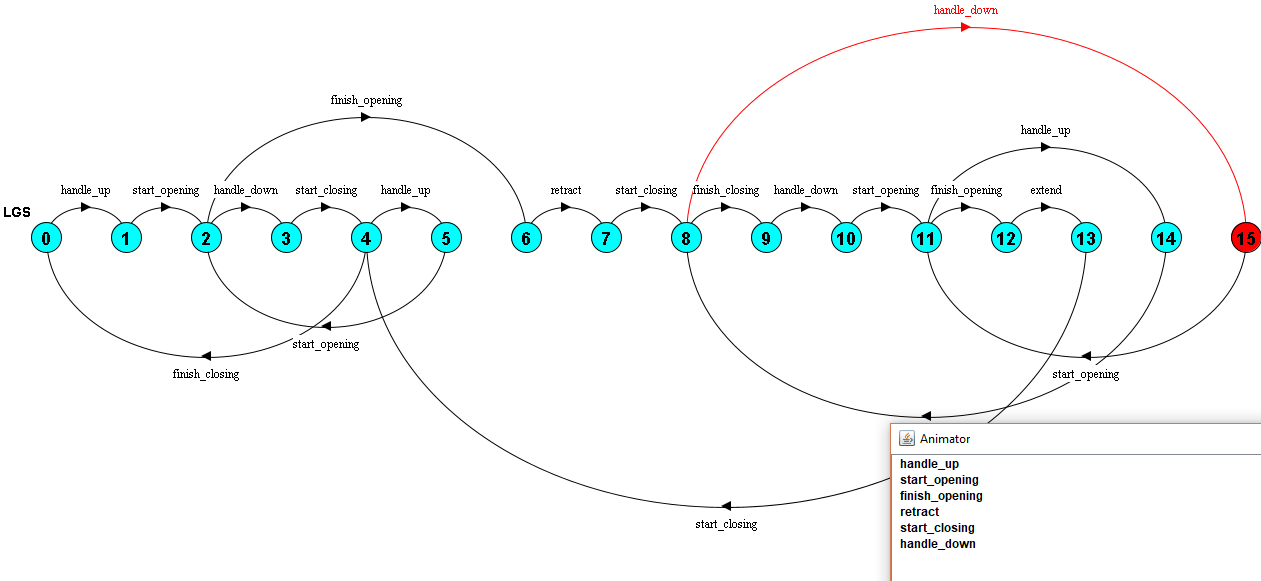}
	\caption{A correctly working LGS state machine.
		Pushing the handle down cancels the gear retraction process and initiates gear extension.
		The bottom-right box contains the trace leading to state 15.}
	\label{fig:correctsm}
\end{figure}
State 15 on \autoref{fig:correctsm} illustrates this situation: the $start\_opening$ outgoing action cancels the door closing process initiated by action $start\_closing$ back in state 7.
The state machine proceeds with the gear extension procedure.
The erroneous ASM specification models a system that waits for the crew to pull the handle up again to let the system complete the gear retraction process.
\begin{figure}
	\centering
	\includegraphics[width=0.9\linewidth]{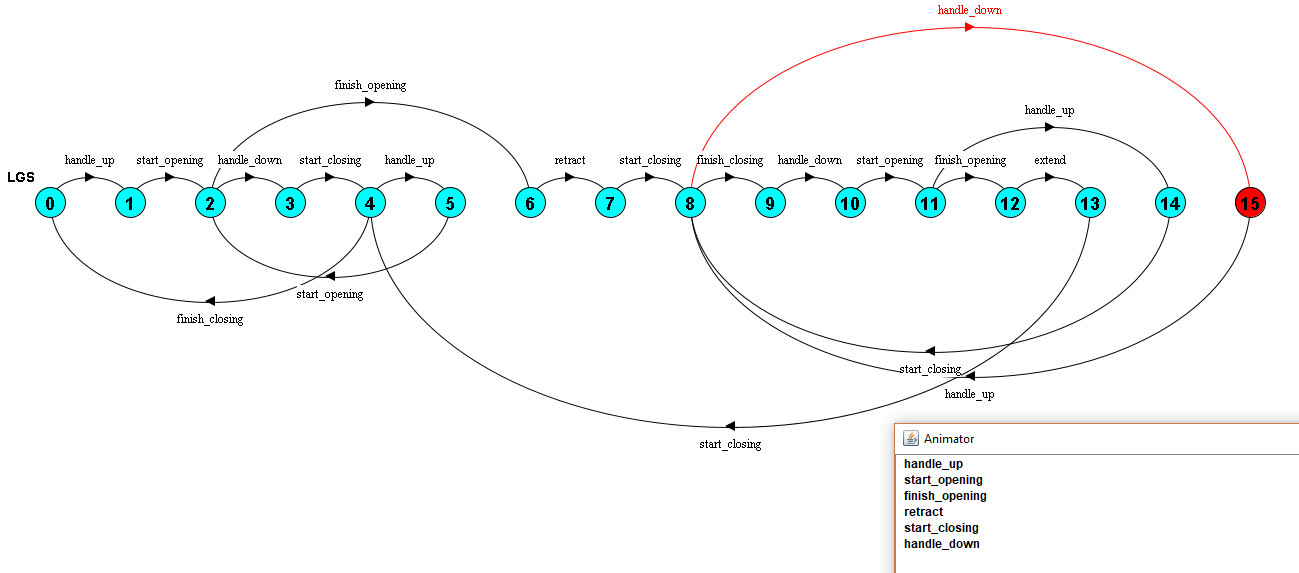}
	\caption{The erroneous LGS state machine.
	Pushing the handle down fails to cancel the gear retraction process.
	It puts the system to waiting for the crew to pull the handle up again.
	The bottom-right box contains the trace leading to state 15.}
	\label{fig:wrongsm}
\end{figure}
State 15 on \autoref{fig:wrongsm} features only one outgoing transition: pulling the handle up again.
Instead of canceling the door closing process (\autoref{fig:correctsm}), the system starts waiting for the crew to pull the handle up.
Imagine a situation in which the crew tries to retract the gear during take-off, and some physical obstacle prevents the latch from closing completely.
In this case a possible solution might be to extend the gear back, and then try to retract it again.
A real controller implemented around the erroneous specification would make extension with the latch partially closed impossible.

The published Eiffel translation of the specification does not have the error.
To catch it with the AutoReq method one needs first to introduce the error back by commenting out two lines in the \textit{open\_door} routine of the Eiffel translation:
\begin{lstlisting}
when closing_state then
door_status := opening_state
\end{lstlisting}
and then submit routine \textit{extension\_duration} to the AutoProof tool; the verification will fail.
The ``README'' file in the accompanying GitHub repository \cite{Naumchev2017LgsRepo} provides detailed instructions on submitting AutoReq requirements to AutoProof.
Internally, AutoProof transforms the Eiffel routine to Boogie code and submits it to the Boogie executable \cite{barnett2005boogie}.
The Boogie executable converts its input to first-order logic formulae and submits them to the Z3 SMT solver \cite{de2008z3}. 

AutoProof detects the error in the following major steps:
\begin{description}
	\item[Inline] the unqualified calls inside of the \textit{extension\_duration} routine to the level of attribute updates and \textbf{\textit{assume}} statements.
	\item[Unroll] the loop inside of \textit{extension\_duration}.
	How much to unroll is a configurable setting; the default configuration suffices for the LGS example.
	\item[Check] the \textbf{\textit{assert}} statements based on the outcome from the \textit{Inline} step.
\end{description}

The intent of applying AutoReq to this example was not to look for errors but to try out the approach, illustrate it on a widely used problem, and compare it with other treatments of that problem.
No error had been reported and we did not expect to find one.
To ascertain its presence, we contacted one of the authors of the original article describing the ASM implementation.
He confirmed the presence of the error in the paper.
(He also noted that the private repository used by his colleagues and him had a correct specification.)


\section{Related work} \label{sec:related_work}

\subsection{Similar studies} \label{sec:sec:similar_studies}

The ASM treatment of the LGS example comes from a collection including other treatments \cite{boniol2014landing}, such as Event-B \cite{su2017aircraft}, \cite{hansen2017validation}, \cite{mammar2017modeling}, Fiacre \cite{berthomieu2014model} and Hybrid Event-B \cite{banach2014landing}.
The original collection \cite{boniol2014landing} discusses pros and cons of these approaches, and the present article does not repeat that discussion.
AutoReq complements these approaches with the following:

\begin{itemize}
\item Language reuse: AutoReq captures temporal and timing properties in a general purpose programming language.
	This will inevitably save resources for software teams that want to apply formal methods.
\item Technology reuse: AutoReq relies on AutoProof, a Hoare logic based program prover.
	The original use case of AutoProof was specifying and verifying programs according to the principles of Design by Contract.
	With AutoReq, software teams can use the tool throughout the whole software lifecycle, starting from the requirements phase.
\item Specification reuse: AutoReq makes it possible to avoid copying-and-pasting already stated assertions through the standard routine call mechanism, familiar to any post-Assembly programmer.
\item Implementation reuse: AutoReq does not require translating programs to models and back for further formal verification. 	
	If a change in the program breaks an AutoReq requirement, the prover will immediately notice this.
\end{itemize}

These advantages need stronger support in the form of successful industrial applications of AutoReq.
Such applications may also uncover additional problems to solve.
The application of AutoReq to the LGS example discussed in the present article inherits the questionable assumptions (\autoref{sec:landing_gear_example}) from the original work by Arcaini et al.
Applying AutoReq to an example with weaker assumptions would provide more evidence of its benefits.

The applicability studies will follow the present article that focuses on illustrating the approach alone.
Combining the first description of AutoReq with its applicability studies would bear the risk of making the article difficult to read.

\subsection{Existing formalisms} \label{sec:sec:existing_limitations}
Reasoning about programs, imperative and concurrent, has been the focus of computer science researchers for decades \cite{jones2003early}, and it traces back as early as Turing's work \cite{jones2017turing}.
Different techniques have been developed over time, and it soon became clear that, while \textit{post facto} verification can be successful for small programs, an effective verification strategy should support and be part of the software development itself and be fully embedded in the process.

The AutoReq method follows this idea and relies on DbC verification; however, one should understand that DbC is not well suited for control systems as it is.
The possibility of unexpected changes in the values of environment-controlled variables introduces the gap between DbC and control systems.
Traditional DbC relies on invariant-based reasoning, on the principle of invariant stability \cite{polikarpova2014flexible}: it should be impossible for an operation to make an object inconsistent without modifying the object.
This principle does not work with control systems because of the unpredictable environment-controlled variables.
In other words, any attempt to constrain an environment-controlled variable through a contract will inevitably lead to the contract's failure.

Control systems communicate asynchronously with the environment.
This introduces another gap with DbC, which is designed from the beginning to deal with synchronous software.
For non-life-critical systems \cite{jackson1995deriving} one may sacrifice the asynchrony under additional assumptions \cite{naumchev2015unifying}, but the Landing Gear System does not fall into this category.

An interesting technique for including environment properties is the notion of monitor introduced by Zave \cite{zave1982operational}.
A monitor is an executable requirement that runs in a dedicated process and observes the system from outside logging possible anomalies.
A monitor continuously polls the state of nondeterministic variables and checks if the system evolves accordingly.
This is, however, a run-time mechanism; in the present work, we seek requirements techniques that lend themselves to static verification.

The general aspiration towards sound static verification resulted in numerous modeling approaches that rely on a declarative logic.
Alloy \cite{jackson2006software} is one of these declarative modeling languages, based on first-order logic, that are used to express complex behavior of software systems.
Alloy is a successor of Z \cite{Abrial:1980} with its own formal syntax and semantics, that adds automatic verification and tool support to Z specifications.
A model created in Alloy can indeed be automatically checked for correctness by using a dedicated tool: the \textit{Alloy Analyzer}, a SAT-based constraint solver that provides fully automatic simulation and checking.
Alloy is one of the tools used for \textit{requirements verification}.
There are several examples of successful applications of the modeling languages in different fields: from pedagogical to enterprise modeling to transportation.
A list documenting some of these applications can be found in \cite{alloyapplications}.

The declarative view simplifies static reasoning, but the system will eventually have to physically operate.
C. A. R. Hoare introduced an imperative logic to statically reason about software way back in 1969.
This invention has been treated as a verification mechanism.
We are interested in requirements specification notations.
The recent notion of seamless requirements \cite{naumchev2017seamless} proposes a use of generalized Hoare triples called specification drivers \cite{naumchev2016complete} as a requirements notation.

The AutoReq method steps forward by applying the idea of seamless requirements to the nondeterministic setting.
It empowers the operational view of Pamela Zave on requirements with AutoProof -- a Hoare logic based prover of Eiffel programs with contracts that relies on the Boogie technology \cite{leino2008boogie}.
In AutoReq a requirement is a routine enriched with \textbf{\textit{assume}} statements capturing environment assumptions and \textbf{\textit{assert}} statements that capture the obligations for AutoProof corresponding to the assumptions.
The resulting method respects environment-controlled phenomena as monitors do but does not assume the requirements to physically run.
The AutoReq method will benefit the development process even when there is no static prover like AutoProof: an operational requirement will become a subject to testing as a parameterized unit test (PUT) \cite{tillmann2005parameterized}.
The testing will consist in this case of running the requirement in the simulated environment described in its \textbf{\textit{assume}} statements.

\subsection{Timing properties} \label{sec:sec:timing}

Modeling real-time computation and related requirements has been a well-investigated matter for long \cite{yamada1962real}.
Representation of real-time requirements, expressed in general or specific form, is a challenging task that has been attacked through several formalisms both in sequential and concurrent settings, and in a broad set of application domains.
The difficulty (or impossibility) to fully represent general real-time requirements other than in natural language or making use of excessively complicated formalisms (unsuitable for software developers), has been recognized.

In \cite{mazzara2010modelling} the domain of real-time reconfiguration of systems is discussed, emphasizing the necessity of adequate formalisms.
The problem of modeling real time in the context of services orchestration in Business Process, and in presence of abnormal behavior has been examined in \cite{mazzara2005timing} and \cite{ferrucci2014ltl} by means, respectively, of process algebra and temporal logic.
Modeling protocols also requires real-time aspects to be represented \cite{berger2000two}.
Event-B has also been used as a vector for real-time extension \cite{iliasov2012augmenting} to handle control systems requirements.

In all these studies, the necessity emerged of focusing on specific typology of requirements using ad-hoc formalisms and techniques and making use of abstractions.
The notion of \textit{real-time} is often abstracted as \textit{number of steps}, a metric commonly used.

The AutoReq method works with the explicit notion of time distance between events by stating operational assumptions on the environment; it also supports the abstraction of time as number of steps through finite loops with integer counters.

\section{Conclusions and future work} \label{sec:conclusions}

The approach presented above is a comprehensive method for requirements analysis based on ideas from modern object-oriented software engineering and the application of a seamless software process that relies on the notation of a programming language as a modeling tool throughout the software process.
The work also introduces the notion of verifying requirements and shows how to use a program prover to perform the verification.
In addition, it connects fundamental concepts, heretofore considered independent, from two different areas of research: verification (\textit{assume}/\textit{assert}) and requirements (\textit{environment}/\textit{machine}).

The work is subject to the following limitations, also suggesting areas of improvement:
\begin{itemize}
	\item While the idea of seamless requirements has been widely applied, its AutoReq development as described here needs more validation on diverse examples in an industrial setting, with actual stakeholders involved.
	\item The patterns given are not necessarily complete; here too experience with more examples is necessary to determine if there is a need for other patterns.
	\item The idea of using a programming language for requirements runs counter to accepted ideas; while there are strong arguments supporting it, and ample discussions in some of the OO literature, some people may still hesitate to adopt it.
	\item More work is required to determine how applicable AutoReq would be to a software process relying on technologies other than Eiffel and AutoProof. In line with this goal, we applied AutoReq \cite{Galinier2018} to the London Ambulance System case \cite{alrajeh2013elaborating}, \cite{letier2001reasoning} and continue working on other examples.

	\item As discussed in \autoref{sec:structuring_an_embedded_system_specification}, parts of the process may benefit from more automation.
	Such further tool support is currently under development.
\end{itemize}

With these reservations, we believe that the article and the case study demonstrate the benefits and contributions listed in the introduction and point to a promising approach to producing and verifying effective requirements for control systems.

\section*{Acknowledgment}
We are indebted to the authors of the ASM version of the LGS case study \cite{arcaini2017rigorous} for their careful work on this problem.
We are particularly grateful to Professor Angelo Gargantini for his openness, patience and insights in discussing the ASM work with us.

The authors are thankful to the administrations of Innopolis University and Toulouse University for the funding that made this work possible.

\newpage
\bibliographystyle{abbrv}
\small\bibliography{library}

\end{document}